\documentclass[12pt,a4paper]{article}
\usepackage{amsfonts}
\usepackage{smart1}
\usepackage{latexsym}

\newcommand{\sectionnew}[1]{\section{#1}}
\link{equation}{section}\toheight1

\newcommand{\beq}{\begin{equation}}
\newcommand{\eeq}{\end{equation}}
\newcommand{\beqa}{\begin{eqnarray}}
\newcommand{\eeqa}{\end{eqnarray}}
\newcommand{\nn}{\nonumber \\}

\newcommand{\spcc}{\vspace{0.2in}}

\def \R {{\mathbb R}}
\def \C {{\mathbb C}}
\def \Z {{\mathbb Z}}
\def \N {{\mathbb N}}
\def \Ss {\mathcal{S}}

\def \f {\left(}
\def \g {\right)}
\def \la {\langle}
\def \ra {\rangle}

\def \a {\backslash}
\def \x {\vec{\xi}}
\def \W {{\cal W}}

\def \M {\overline{M}}
\def \MM {\overline{\overline{M}}}
\def \l {\lambda}
\def \X {\mathrm{x}} 
\def \mg {M^{\times n}_{\f g \g }}
\def \Wgg {\W_{\f g \g} \f x_1,...,x_n \g }
\def \Wx {\W \f x_1,...,x_n \g }

\advance\topmargin-2cm

\begin{document}


\title{
Rationality of conformally invariant
local correlation functions on
compactified Minkowski space
}

\author{
Nikolay M. Nikolov\footnote{e-mail: mitov@inrne.bas.bg}
\quad and \quad
Ivan T. Todorov\footnote{e-mail: todorov@inrne.bas.bg}
\\
\\
{\normalsize
Institute for Nuclear Research and Nuclear Energy
}
\\
{\normalsize
72 Tsarigradsko Chaussee, BG-1784 Sofia, Bulgaria
}}

\date{October \ 23, \ 2000}
\maketitle


\begin{abstract}


Rationality of the Wightman functions
is proven to follow from 
energy positivity, locality and a natural condition
of global conformal invariance (GCI)
in any number $D$ of space-time dimensions.
The GCI condition allows to treat correlation functions
as generalized sections
of a vector bundle
over the compactification
$\M$ of Minkowski space $M$ and yields a strong form of
locality valid for all non-isotropic intervals if
assumed true for space-like separations.

\end{abstract}




\noindent
{\bf\Large Introduction} \\

The study of conformal (quantum)
field theory (CFT) in four (or,
in fact, any number of) space time dimensions (see, e.g.,
\cite{Di 36} \cite{W 60} \cite{Po 70} \cite{MS 72}
\cite{MT 73} \cite{Po 74} \cite{LM 75} \cite{M 77'}
\cite{M 77} \cite{RY 77} \cite{Stanev 88}
as well as the reviews
\cite{TMP 78} \cite{To 85} \cite{To 86}
and references therein)
preceded the continuing excitement with
$2$-dimensional (2D) CFT
(for a modern textbook and references
to original work - see \cite{DMS 96}).
Interest in higher dimensional
CFT was revived (starting in late 1997)
by the discovery of the AdS-CFT
correspondence in the context of string
theory and supergravity
(recent advances in this crowded field
can be traced back from
\cite{AGMOL 99}).
The present
paper is more conservative in scope:
we try to revitalize the old
program of combining conformal
invariance and operator product
expansion with the general
principles of quantum field theory
using some insight gained in the study of 2D CFT models.

The main result of the paper (Theorems 3.1 and 4.1) can
be formulated (omitting technicalities) as follows.

We say that a QFT (obeying Wightman axioms \cite{SW 78}, \cite{Jost})
satisfies \textit{global conformal invariance} (GCI) if
for any conformal transformation $g$ and for any set of
points $\f x_1,...,x_n \g$ in Minkowski space
$M$ such that their images
$\f g \, x_1 \, ,\, ...\, ,\, g \, x_n \g$
also lie in $M$ the Wightman function $\W \f x_1,...,x_n \g$
stays invariant under $g \,$.
We note that this requirement (stated more precisely in
Sec.2) is stronger than the one used (under the same name)
in \cite{LM 75}.
An important tool in using GCI is the fact that mutually
non-isotropic pairs of points form a single conformal
orbit on compactified Minkowski space (Proposition 1.1).
Together with local commutativity for space like separations
this implies the vanishing of the commutator
$\left[ \, \phi_1 \f x_1 \g \, , \,
\phi_2 \f x_2 \g \, \right]$
whenever the difference
$x_{12} = x_1 - x_2$ is non-isotropic
(it follows from Lemma 3.2).
We deduce from this strong locality property
combined with the energy positivity condition
that the Wightman functions are rational
in $x_{ij}$ (Theorem 3.1).
Hilbert space positivity gives strong restrictions on the
degrees of the poles of the resulting rational functions
(Theorems 4.1, 4.2 and Proposition 4.3).

This result severely limits the class of local QFT
satisfying GCI and makes feasible the construction
of general conformal invariant correlation functions in
such theories.
That is illustrated on a simple example in the concluding
Sec. 5.
The interesting problem of
exploiting the presence of conserved currents and
the stress energy tensor and classifying their
correlation functions in a QFT satisfying
GCI in $4$ space-time dimensions is the subject of
a separate study carried out currently in collaboration
with Yassen Stanev.




\section{
Conformally compactified Minkowski space $\M$
and its complexification. The orbit of non-isotropic
pairs of points in $\M \times \M$
}

\textit{Conformally compactified Minkowski space}
$\M$ of dimension
$D$ is a homogeneous space of the
\textit{connected} conformal group
$C_0$ (for even $D$, $C_0$ is
isomorphic to $SO_0 \f D,2 \g / \Z_2$;
for odd $D\;$, $C_0 \cong SO_0 \f D,2 \g \,$).
Unless otherwise stated we shall
suppose that the dimension
is an arbitrary integer $D \geq 1\,$.
We shall use also in the exceptional cases $D=1, \ 2$
the group $SO_0 \f D,2 \g$ of M\"{o}bius transformations
(as the infinite dimensional conformal group present in
those cases is \textit{not} an invariance group of
the vacuum state).
The \textit{Minkowski space} $M$ is embedded
as a dense open
subset in $\M$ in such a way that
the isotropy relation in $M \times M$
extends to a conformally invariant "isotropy relation"
in $\M \times \M \,$.
More generally, for every element
$g \in SO_0 \f D,2 \g$ there exists a quadratic polynomial
$\omega \f x , g \g$ in the coordinates
$x$ in $M \cong \R^{D-1,1}$
such that the pseudo-euclidean interval transforms as:
\beq\label{1:1}
{\f gx-gy\g}^2\ =\
\frac{{\f x-y\g}^2}{\omega\f x,g \g\omega\f y,g \g}
\quad , \qquad
\eeq
where $x \mapsto gx$ is the
\textit{nonlinear} (coordinate)
conformal action of $SO_0 \f D,2 \g$
on $M$ (with singularities, see Appendix A). The complement
$K_{\infty}:=\M \a M$, the set of
\textit{points at infinity}, is an
isotropic $(D-1)$-cone: there is a
unique point $p_{\infty} \in K_{\infty} \,$,
the tip of the cone,
such that $K_{\infty}$ is the set of
all points $p$ in $\M$ isotropic
to $p_{\infty}$. Thus the
\textit{stabilizer} $C_{\infty}$ of
$p_{\infty}$ in $C_0$
leaves $M$ (and $K_{\infty} \,$) invariant;
it is the
\textit{Poincar\'{e} group with dilations} of $M$
(also called the Weyl group).

\textbf{Proposition 1.1}  \textit{Any pair} $(p_0,p_1)$
\textit{of mutually non-isotropic points of} $\M$
\textit{can be mapped into any other such pair}
$(p'_0,p'_1)$ \textit{by a conformal transformation}.

\textit{Proof.} Due to the transitivity of the action of $C_0$
there are elements $g_0$ and $g'_0$ of $C_0$ which carry
$p_0$ and $p'_0$ into the point $p_{\infty} \,$:
$g_0 p_0 = g'_0 p'_0 = p_{\infty} \,$.
Then
the images $g_0 p_1$ and $g'_0 p'_1$ of the two other points
will both
belong to $M$
(because the original pairs are mutually non isotropic)
and can hence be moved to one another by a
translation $t$ (in $C_{\infty} \,$) which leaves
$p_{\infty}$ invariant: $g'_0 p'_1 = t g_0 p_1 \,$.
So
the element $g \in C_0$ which transforms the pair
$\f p_0 , p_1 \g$ into $\f p'_0 , p'_1 \g$ is
given by
$g= {\f g'_0 \g }^{-1} t \, g_0 \,$.$\quad\Box$
\spcc

\textit{Remark 1.1} It is important
to note that we are just dealing in
Proposition 1.1 with pairs of points: a continuous
time-like world line cannot be mapped into a space-like
one by a conformal transformation. In fact, the connected
conformal group preserves causal ordering on such lines-
see \cite{FS 66} and \cite{Se 71}.$\quad\Box$
\spcc

We shall be primarily interested
in this paper in the space-time bundle
of (Fermi and) Bose fields of
(half)integer dimension transforming under
a representation of the finite
covering $C=Spin_0 \f D,2 \g$ of the
conformal group $C_0$ of $\M$.
(For even $D$, $C$ is a $4$-fold
covering of $C_0=SO_0 \f D,2 \g / \Z_2$;
for odd $D$ it is a $2$-fold
covering of $C_0=SO_0 \f D,2 \g $.)
The group $C$ acts (\textit{transitively})
on $\M$ through its canonical
projection on $C_0$ (its centre acting trivially).

We shall use
the following atlas on
$\M \,$.
Let $x : M \cong \R^{D-1,1}$ be
the standard Minkowski chart in $\M \,$;
for every $g \in C$ we set
$M_{\f g \g }:=g^{-1}M$ 
and introduce the coordinatization map
$x_{\f g \g }: = x \circ g : M_{\f g \g } \cong \R^{D-1,1}\,$
thus obtaining an atlas
$\left\{x_{\f g \g }\, ; \, g \in C\right\}$ over $\M\,$.
The transformation from the
coordinates $x_{\f g \g }$  to $x_{\f g' \g }$
is $x_{\f g' \g }=g'g^{-1}x_{\f g \g }\,$.
(One can in fact show that $\M$ can be covered by just
$3$ charts of this type.)
We need also an atlas on $\M^{\times n}$ in the study
of
$n$-point correlation functions.
To this end we note that such an atlas is
given by the diagonal subsystem of the Cartesian power
of the above atlas:
\beq\label{I.1}
\left\{\
\f x_{1 \, \f g \g },...,x_{n \, \f g \g } \g
\; : \; \mg \cong \R^{D \,n} \quad ; \quad g \in C
\right\} \qquad
\eeq
(to simplify notation we write here and in what
follows $\R^D$ instead of $\R^{D-1,1} \,$).
Indeed,
$\mathop{\bigcup} \limits_{g \in C} \mg= \M^{\times n}\,$,
which is equivalent to the statement
that for every set of points
$p_1,...,p_n \in \M$ there exists $g \in C$ such that
$gp_1,...,gp_n \in M\ $. The last statement
can be proven by induction in $n$
(see also the argument in Appendix C
and the proof of Lemma 3.2).

Finally, let us introduce
the complexification $\M_{\C}$ of $\M \,$.
It is needed
because of the condition of \textit{energy positivity}
in QFT, which implies that the vector valued distribution
$F(x)=\phi (x)|0\ra$ where $\phi (x)$ is an arbitrary (local)
Wightman field \cite{SW 78}, \cite{Jost}
can be viewed as the boundary value of
an analytic function $F(x+iy)$ holomorphic in the (forward)
\textit{tube domain} $T^+$ where
\beq\label{1.8}
T^{\pm}=\
\left\{x\pm iy\ ;\ x\in M\ ,\ y\in V_+ \right\}\, ,\quad
V_{\pm} =\
\left\{y\in M\ ;\ \pm y^0>|\underline{y}|\right\}.\quad
\eeq
Clearly, $T^{\pm}\subset \M_\C$ and each of them is a
\textit{homogeneous space}
of the (real) conformal group $C$ \cite{Uhl 63},
the stabilizer of a point being conjugate to the
maximal compact subgroup
$Spin \f D \g \times Spin \f 2 \g$ of $C \,$.



\sectionnew{
Global conformal invariance of Wightman functions
}

The Wightman functions,
the vacuum expectation values of products 
$\phi_1 \f x_1 \g ... \phi_n \f x_n \g$
of a multicomponent field
$\phi \f x \g \,$,
are defined
as tensor valued
tempered distributions on
$M^{\times n} \cong \R^{D \, n} \,$:
\beq\label{2|1}
{\cal W}_{1...n}\left(x_1,...,x_n\right)\ =\ \la 0|
\ \phi_1\f x_1\g\phi_2\f x_2\g ...\phi_n\f x_n\g |0\ra\ \in\
\Ss '\left(M^{\times n},\ F^{\otimes n}\right).\quad
\eeq
where $F$ is a finite dimensional (complex) vector space.
In Eq.(\ref{2|1}) and further we are using
Faddeev's shorthand for tensor products
($\phi_1 \, ... \, \phi_n =
\phi \f x_1 \g \otimes ... \otimes \phi \f x_n \g \,$).

We shall assume that
$C$
acts (locally) on
$M\times F\ $:
\beqa\label{2|2}
&&M\times F \ \ni\ \left(x,v\right)
\ \mathop{\longmapsto}\limits^{g}
\ \left(gx,\pi_x\left(g\right)v\right) \ \in \ M\times F
\quad \mathrm{iff} \quad gx \in M \qquad\qquad\qquad
\nn
&& \quad \mathrm{where} \quad \pi_x\left(g\right)
\ \in\ Aut\f F \g \quad ,
\quad \pi_x\left(g_1g_2\right)=
\pi_{g_2x}\left(g_1\right)\pi_x
\left(g_2\right) \qquad
\eeqa
$\pi_x\f g\g$
being a
\textit{single-valued}
real analytic function
defined for
all pairs
$x\in M \,$, $g\in C$
for which $gx\in M$
(thus the multiplicativity of
$\pi_x$ in (\ref{2|2})
holds under the provision
that both
$g_2x$ and $g_1g_2x$ belong to $M\,$).
Moreover, we will assume that if
$g$
is a translation,
$t_a x =x+a\,$,
then
$\pi_x$
acts trivially on
$F$, -i.e., $\pi_x\f t_a\g =id\,$.
Note that these conditions are satisfied
by the usually considered
local induced representations of
$C$
(see \cite{M 77} or Chapter 2 of \cite{To 85}).

\textit{Example 2.1} Let $\phi$ be a vector \textit{current}
$j^{\mu}\f x\g$; then $dim\ F=D$. For $x^2\neq 0$
the action of
the \textit{Weyl reflection} $w \, \f \in C_0 \g$ is given by:
\beq\label{2||6}
w:\f x^{\mu},j^{\mu}\g \ \mapsto \ \f \frac{x^0}{x^2}\ ,\
\frac{-\underline{x}}{x^2}
\ ;\ -\frac{r_{\mu\nu}\f x\g}{{\f x^2\g}^d}
\ j^{\nu}\g \quad ,\quad
{r^{\mu}}_{\nu}\f x\g={\delta^{\mu}_{\nu}} -
2\ \frac{x^{\mu}x_{\nu}}{x^2} \ ,\qquad
\eeq
where $d$ is the conformal dimension of the current $j\,$.
The result
differs by a space reflection from the
\textit{conformal inversion},
$I_r :\f x^{\mu},j^{\mu}\g
\ \mapsto \ \f \frac{x^{\mu}}{x^2}\ ,\
\frac{{r^{\mu}}_{\nu}\f x\g}{{\f x^2\g}^d} j^{\nu}\g$ (which
does not belong to $C_0 \,$).
It is easy to verify that $I_r$,
and hence $w$, leave invariant
the current conservation law iff $d=D-1$;
indeed, this follows from the
identity
\beq
\partial_{\mu}\f \frac{{r^{\mu}}_{\nu}\f x\g}{{\f x^2\g}^d}
\ j^{\nu}\f \frac{x}{x^2}\g \g\ =\
\frac{1}{{\f x^2\g}^{d+1}}\ \partial_{\nu} j^{\nu}
\f \frac{x}{x^2}\g
+2\ \frac{d+1-D}{{\f x^2\g}^{d+1}}\ x_{\nu}\ j^{\nu}
\f \frac{x}{x^2}\g\ .\quad
\eeq
{\samepage
Note that the function $\pi_x\f w\g$
(as well as $\pi_x \f I_r\g$)
is well defined and single-valued (for all $x$ such that
$x^2\neq 0$), iff $d$ \textit{is an integer}.$\quad\Box$
\spcc
}

{\samepage
\textit{Remark 2.1} The condition for a
trivial action of the translations in
$M$ onto $F\ $- i.e. $\pi_x\f t_a\g =id\ $-
is not restrictive.
If it is not
satisfied we can pass to an equivalent action:
\beq\label{2|3}
\pi'_x\f g\g
\ :=\ \pi_{gx}\f t^{-1}_{gx}\g \pi_x\f g\g \pi_0\f t_x\g
\ =\ \pi_0\f t^{-1}_{gx}gt_x\g \qquad
\eeq
for which
$\pi'_x\f t_a\g = id$ does hold.
It also follows from (\ref{2|3}) that
the growth of
$\pi_x\f g\g$
for $x \to \infty$
is not more than polynomial.$\quad\Box$
\spcc
}

The fact that the map (\ref{2|2}) is single valued
outside its singularities allows to treat
$\pi_x\f g\g$
as a cocycle on a fibre bundle over
$\M$
with a standard fibre
$F\ $.
More generally, for every
$n=1,2,3,...$
and the atlas (\ref{I.1}) we have a cocycle:
\beqa\label{2||3}
&& \f \X_{\f g \g },\W \g \ \longmapsto \
\f \,\X_{\f g' \g }\ =\ g'g^{-1}\X_{\f g \g }\, ,\
\pi_{\f n\g} \f \X_{\f g \g } \, ; \, g'g^{-1} \g \W \,\g
\;\; \in \; M^{\times n} \times F^{\otimes n}
\qquad\qquad\qquad
\nn
&&\mathrm{where:} \nn
&&g\X\ :=\ \f gx_1,...,gx_n \g
\quad ;\quad
\pi_{\f n \g} \f \X \, ; \, g \g\ :=\
\pi_{x_1}\f g \g \otimes\, ...\, \otimes\pi_{x_n}\f g \g
\quad \nn &&
\mathrm{for} \quad \X\ =\ \f x_1,...,x_n \g
\quad \mathrm{and} \quad
g\in C \qquad
\eeqa
which gives a fibre bundle $E_n$ over
$\M^{\times n}$
with a standard fibre
$F^{\otimes n}\ $.
It allows to consider the Wightman functions $\W_{1...n}$
as generalized sections $\W$ of
$E_n$
which admit for each chart $\mg$ of the
atlas (\ref{I.1}) a coordinate expression
$\Wgg \in {\cal D}' \f \R^{Dn},F^{\otimes n} \g \,$
(${\cal D}'$ being the space of
distributions over
test functions of compact support).
These coordinate
expressions should satisfy the consistency condition
\beqa\label{N2.6}
{\cal W}_{\f g' \g} \f g'g^{-1}x_1,...,g'g^{-1}x_n \g
\ = \
\pi_{\f n \g} \f \X \, ; \, g'g^{-1} \g
\,
{\cal W}_{\f g \g} \f x_1,...,x_n \g
\quad \qquad
\eeqa
(with the cocycle (\ref{2||3})).
We note that Eq. (\ref{N2.6})
is understood locally around every
$\X \in \R^{D \, n}$
such  that $g'g^{-1}\X \in \R^{D \, n} \,$:
then the transformation of
the coordinates $\X \mapsto g'g^{-1} \X$
is a diffeomorphism and
$\pi_{\f n \g} \f \X \, ; \, g'g^{-1} \g$
is a multiplicator.
We can further define
the space ${\cal D}_n$ of test sections for a generalized
section $\W \,$, which is actually the space of all smooth
sections of the fibre bundle over $\M^{\times n}$ with a
standard fibre $F^{* \ \otimes n}$ and the cocycle:
\beqa\label{2|4}
&& \f \X_{\f g \g },f \g \ \longmapsto \
\f \,\X_{\f g' \g }\ =\ g'g^{-1}\X_{\f g \g }\, ,\
\pi^{\f n\g} \f \X_{\f g \g } \, ; \, g'g^{-1} \g f \,\g
\:\; \in \; M^{\times n} \times F^{* \ \otimes n}
\qquad\qquad\qquad
\nn
&&\mathrm{where:} \nn
&&\pi^{\f n \g} \f \X \, ; \, g \g\ :=\
{\pi_{\f n \g} \f \X \, ; \, g \g }^{-1\, *} \
{J\f \X \, ; \, g \g }^{-1}
\quad \qquad
\eeqa
and
$J$ is the Jacobian of the transformation $x \mapsto gx \,$.
(For the concepts of test functions and distributions on a
manifold see, also, \cite{Treves}.)

{\samepage
{\bf Proposition 2.1} (\textit{a}) \textit{Let the
distributions} $\W_{\f g \g}$ \textit{be defined for any}
$g \in C$ \textit{and satisfy the consistency condition}
(\ref{N2.6}). \textit{Then there exists a unique linear
functional} $\W$ \textit{on} ${\cal D}_n$ \textit{that
is a generalized section of the vector bundle} $E_n$
\textit{with coordinate expressions} $\W_{\f g \g} \,$.

(\textit{b}) \textit{Each} $\W_{\f g \g}$ \textit{actually
belongs to the subspace of tempered distributions}
${\cal S}' \f \R^{Dn},F^{\otimes n} \g$ \textit{of}
${\cal D}' \f \R^{Dn},F^{\otimes n} \g \,$.
}

\textit{Sketch of proof.} (\textit{a}) Since $\M^{\times n}$
is compact there exists a finite partition of unity
$1 = \mathop{\sum}\limits_{i=1}^k a_i \,$, where $a_i$ is a
smooth function of compact support in
some charts
$\M_{\f g_i \g}^{\; \times n}\,$
($i=1,...,k$).
The linear functional $\W$
is then
uniquely
defined by
\beq\label{N2.7}
\la \, \W , f \, \ra \ :=\
\mathop{\sum} \limits_{i=1}^{k} \
\la \ \W_{\f g_i \g } \, , \,
{\f a_i f \g }_{\f g_i \g } \f \X \g \ \ra
\quad ,\qquad
\eeq
where ${\f a_i f \g }_{\f g_i \g } \f \X \g$
is the coordinate
expression of the test section $a_i f \,$.

{\samepage
(\textit{b}) Getting a test section  $f$ with a coordinate
expression 
$f_{\f g \g} \f \X \g \in
{\cal S} \f \R^{Dn},F^{\otimes n} \g \,$,
we obtain this statement by Eq. (\ref{N2.7}) because of the
continuity of the linear maps:
\beq\label{2|11}
{\cal S} \f \R^{Dn},F^{\otimes n} \g
\;\; \ni \;\; \ f_{\f g \g} \f \X \g \ \longmapsto \
{\f a_if \g }_{\f g_i \g } \f \X \g
\;\; \in \;\; {\cal D} \f \R^{Dn},F^{\otimes n} \g
\quad . \qquad
\eeq
$\Box$
\spcc
}

The action (\ref{2|2}) not only defines a cocycle on $E_n\,$;
it also gives rise to a natural action of the conformal group
$C$ on $E_n$ that is linear on the fibres. \textit{A
generalized section} $\W$ of $E_n$ is called
\textit{conformally invariant} if it does not change under
the action of $C \,$. This is equivalent to requiring that
$\W$ has the same coordinate expression in every chart:
\beqa\label{N2.9}
\W_{\f g \g} \f x_1,...,x_n \g \ = \
\W \f x_1,...,x_n \g \quad . \qquad
\eeqa
Because of the consistency condition (\ref{N2.6}) this is
equivalent to the following requirement.

(\textit{GCI}) \textit{Global conformal invariance}
\beq\label{2.7} 
{\pi_{x_1}\f g \g }^{-1} \ \otimes \ .\ .\ .\ \otimes \
{\pi_{x_n}\f g \g }^{-1} \
\W\f gx_1,...,gx_n \g\ =\ \W\f x_1,...,x_n \g
\quad
\mathrm{for} \quad gx_1,...,gx_n \in M \quad .\qquad
\eeq 
This implies, in particular,
(unrestricted) Poincar\'{e} and dilation
invariance. We recall that
Eq.(\ref{2.7}) should be interpreted locally: for a
fixed $g\in C\,$,
${\pi_{x_1}\f g \g }^{-1}
\otimes ... \otimes {\pi_{x_n}\f g \g }^{-1}$
should be viewed as a multiplicator
(regular in the neighbourhood of
$\f x_1,...,x_n\g\in M^{\times n}$) while
$g:\f x_1,...,x_n\g\mapsto \f gx_1,...,gx_n\g$
is a local diffeomorphism.
(That is, both sides of (\ref{2.7})
are to be smeared with test
functions with support in the above neighbourhood.)
\spcc

Thus we have an one-to-one correspondence
between the tempered
distributions \\
$\W \f x_1,...,x_n \g \in
{\cal S}' \f \R^{Dn},F^{\otimes n} \g$
satisfying the condition (\textit{GCI})
and the conformal invariant
generalized sections of the bundle $E_n \,$.

\textit{Remark 2.2} \textit{Local conformal invariance}
requires the
existence of a continuous curve $g\f \tau\g\in C$, such that
$g\f 0\g=1$, $g\f 1\g=g$ and
$\f g\f \tau\g x_1,...,g\f \tau\g x_n\g \in M^{\times n}$
for all
$\tau \in [0,1]$; it is equivalent to
invariance under infinitesimal
conformal transformations.
The (\textit{GCI}) condition (\ref{2.7})
does not demand the existence of such a curve;
moreover, if several
curves of this type exist it says that
$\pi_x\f g\g$ is independent
of the path $g\f \tau\g$ connecting $g$
with the group unit. The
$2$-point function of a hermitean scalar field
of dimension $d$,
\beqa\label{2:10}
W_d\f x_{12}\g
\ =\ \frac{\Gamma(d)}{{\f 4\pi\g}^{\frac{D}{2}}}
\ {\f\frac{4}{x_{12}^{\ 2}+i0x_{12}^0}\g}^d\ =\
\frac{2\pi}{\Gamma\f d+1-\frac{D}{2}\g}\
\mathop{\int} \theta \f p^0\g
\frac{\theta \f -p^2\g e^{ipx}}{{\f -p^2\g}^{\frac{D}{2}-d}}
\ \frac{d^D p}{{\f 2\pi\g}^D}\quad  \qquad
\eeqa
($\, x_{12} = x_1 - x_2 \,$),
satisfies all Wightman axioms (including positivity for any
$d\geq \frac{D}{2}-1$, $D>2$) as well as local conformal
invariance, however, it only obeys
(\textit{GCI}) for positive
integer $d$, because of the
requirement of singlevaluedness of $\pi_x\f g\g$ inherent in
(\textit{GCI}). Thus (\textit{GCI}) is indeed stronger than
local conformal invariance.$\quad\Box$
\spcc

The invariance of the $i0x_{12}^0$
prescription in (\ref{2:10})
is implied by the following more general statement.

{\bf Proposition 2.2} \textit{Any product of}
$2$-\textit{point like
functions of type} (\ref{2:10})
\textit{satisfies the invariance
condition} (\textit{GCI}):
\beqa\label{2:11}
&&
\left[\omega\f x_1,g \g \right]^{-\mu_1}\ ...\
\left[\omega\f x_n,g \g \right]^{-\mu_n}
\mathop{\prod}\limits_{1\leq j<k\leq n}
{\left[ {\f gx_j -gx_k \g}^{\ 2}+i0\f gx_j^0 -gx_k^0 \g
\right]}^{-\mu_{jk}}
\ = \nn
&&
=\ \mathop{\prod}\limits_{1\leq j<k\leq n}
{\f x_{jk}^{\ 2}+i0x_{jk}^0 \g }^{-\mu_{jk}}\; ,
\qquad
\eeqa
\textit{where} $\mu_{jk} \in \Z\ $,
$\mu_k=\sum_{j=1}^{k-1}\ \mu_{jk}
\ +\  \sum_{l=k+1}^{n}\ \mu_{kl}\ $,
\textit{and the factors}
$\omega \f x,g \g$ \textit{are introduced in}
(\ref{1:1}).

Away from the singularities, for
$x_{ij}^{\ 2}\neq 0\ ,$ the statement
follows from Eq. (\ref{1:1}).
Only establishing
conformal invariance of the
"$+i0x_{ij}^0 \ "$ prescription
poses a problem.
The somewhat technical \textit{proof} of this
statement is relegated to
Appendix B. It is based on the fact that
${\f x_{jk}^{\ 2}+i0x_{jk}^0 \g }^{-\mu}$
is a limit of a holomorphic
function of $x_j-\zeta_k$ for $T_+ \ni \zeta_k \to x_k$
and of the conformal invariance of the forward tube
(in the transformation
of $\zeta_k$ in
$x_j-\zeta_k\ \mapsto \ gx_j - g\zeta_k \ $).$\quad\Box$
\spcc

{\samepage
\textit{Example 2.2} Let the  points
$x_1$, $x_2$ of $M$ satisfy
$x_{12}^{\ 2}>0$ and $x_1^{\ 2}\ x_2^{\ 2}<0$.
Then $wx_1,wx_2 \in M$
(where $wx$ is defined in Example 2.1)
but there is no continuous curve
$g\f \tau\g\in C$  that relates $\f x_1,x_2\g$ with $\f wx_1,wx_2\g$
in $M\times M$. Indeed,
${\f wx_1-wx_2\g}^2=
\frac{x_{12}^{\ 2}}{x_1^{\ 2}\ x_2^{\ 2}}<0$; if there
were $x_i\f \tau\g = g\f \tau\g x_i \in M$, $i=1,2$
which connect
continuously $x_i$ with $wx_i$
then there would have been a point
$0<\tau_0<1$ such that
${\f x_1\f \tau_0\g -x_2\f \tau_0\g\g}^2=0$
which is impossible (cf. Remark 1.1).$\quad\Box$}



\sectionnew{
Wightman axioms and
global conformal invariance imply rationality
}

We will recall first some general properties of Wightman
functions (\ref{2|1}) (see for more detail \cite{SW 78} or \cite{Jost}).
Thus $\cal W$ is a
tempered distribution with
values in the $n$-fold tensor power
$F^{\otimes n}$ of a finite
dimensional (complex) vector space $F\ $.
For complex fields the vector $\phi\in F$
is assumed to contain
along with each component $\phi_a$ also its
\textit{conjugate}
$\phi_{\overline{a}}={\phi_a}^*$.
The splitting of the fields
into bosonic and fermionic ones
amounts to defining a $*$-invariant
$\Z_2$-\textit{grading}
$F=F_0\oplus F_1\ .$ Then the automorphisms
$\pi_x \f g \g$ of $F$
in the conformal action (\ref{2|2}) will
preserve the conjugation and the $\Z_2$-grading of $F\ $.
The implications of translation
invariance and energy positivity and of locality can be
summed up in the
following conditions for $\Wx \ $.

(\textit{TS}) \textit{Translation invariance and spectral
condition. The Fourier transform of}
\beq\label{2.2}
{\cal W}\left(x_1,...,x_n\right)\ =\ W\left(
x_1-x_2\ ,\ ...\ ,\ x_{n-1}-x_n\right)\quad
\eeq
(omitting the indices $1...n$ in both sides),
\beq\label{2.3} 
\tilde{W} \left(q_1,...,q_{n-1}\right)\ =\
\mathop{\int \cdots \int} \limits_{M^{\times (n-1)}}
W \left( y_1,...,y_{n-1} \right)
\ e^{-i\left( q_1y_1+...+q_{n-1}y_{n-1} \right)}
\ d^Dy_1...d^Dy_{n-1}
\quad ,
\qquad
\eeq
{\samepage
has support in the product of
(closed) future light cones in $M$:
\beq\label{2.4}
supp\ \tilde{W}\subseteq
\f {\overline{V}}^{\ +}\g^{\times (n-1)}
\quad ,\quad
{\overline{V}}^{\ \pm}=\left\{q\in M\ ;\ \pm q^0 \geq
|\underline{q}| \right\}.\qquad
\eeq

This is the relativistic (Lorentz invariant) form of
\textit{energy positivity}.
\spcc

(\textit{L}) \textit{Locality}
\beqa\label{Loc}
\W_{1\ ...\ i\ i+1\ ...\ n}
\left(x_1,...,x_i,x_{i+1},...,x_n\right) && =\;\;
\epsilon_{i\ i+1}\ \W_{1\ ...\ i+1\ i\ ...\ n}
\left(x_1,...,x_{i+1},x_i,...,x_n\right)
\qquad \qquad \nn &&
\mathrm{for} \quad {x_{i\ i+1}}^2>0 \qquad \qquad
\eeqa
($x_{i\ i+1}=x_i-x_{i+1}$);
where $\epsilon_{ij}$ $\f i\neq j \g$ is a sign factor,
$\epsilon_{ij}=-1$ if both $i$ and $j$
refer to Fermi fields (elements of $F_1$)
and $\epsilon_{ij}=1$ otherwise.
In other words
Fermi fields anticommute among themselves while,
Bose fields commute
with both Bose and Fermi fields
for space-like separations.}
\spcc

Permutation of field indices accompanied by the sign factors
$\epsilon_{ij}$ give rise to an action of
the symmetic group $S_n$
on $F^{\otimes n}$. An $n$-point function
${\cal F} : M^{\times n} \longrightarrow F^{\otimes n}$
is said to be
\textit{$\Z_2$-symmetric}
if it is invariant under this action
combined with the corresponding permutations of the
coordinates.

\textit{Remark 3.1}  The restriction (\ref{2|1})
on the class of
distributions is not essential
for theories satisfying (\textit{GCI}).
Indeed, if we assume $\W \in {\cal D}'$
(or $\tilde{\W} \in {\cal D}'$)
then dilation invariance
(which has a similar form in coordinate and
in momentum space) implies
$\W \in {\Ss}'$ (\ref{2|1}).$\quad\Box$
\spcc

{\bf Theorem 3.1} \textit{The tempered distribution}
$\W_{1...n}\f x_1,...,x_n\g$ \textit{satisfies
conditions} (\textit{TS}), (\textit{L})
\textit{and} (\textit{GCI})
\textit{iff it can be expressed in terms of
a rational function of the type}
\beqa\label{2:12}
&&
\W_{1...n}\f x_1,...,x_n\g
\ =\ {\cal P}_{1...n}\f x_1,...,x_n\g
\mathop{\prod}\limits_{1\leq j<k\leq n}
{\f x_{jk}^{\ 2}+i0x_{jk}^0 \g }^{-\mu^n_{jk}} \quad ,\qquad
\nn &&
\qquad \qquad \quad \;\  \f \; x_{jk} = x_j - x_k \; \g \quad .\qquad
\eeqa
\textit{Here} $\mu^n_{jk} \geq 0$ \textit{are integers},
${\cal P}_{1...n}\f x_1,...,x_n\g$
\textit{is a polynomial with values in}
$F^{ \otimes n}$
\textit{and the associated rational function}
${\cal R}_{1...n}\f x_1,...,x_n\g
\ =\ {\cal P}_{1...n}\f x_1,...,x_n\g
\mathop{\prod}\limits_{1\leq j<k\leq n}
{\f x_{jk}^{\ 2} \g}^{-\mu^n_{jk}}$
\textit{is fully $\Z_2$-symmetric
and satisfies the conformal invariance condition} (\ref{2.7})
\textit{as a rational function}.

The \textit{proof} of this Theorem is based on the following

{\bf Lemma 3.2} \textit{For each set of points}
$\f x_1,...,x_m,y_1,y_2\g$ \textit{in}
$M$ \textit{such that}
$y_{12}^{\ 2}\neq 0$
\textit{and a pair of mutually non-isotropic}
$y_1'$, $y_2'$ \textit{there exists a}
$g\in C$ \textit{such that}
$gx_i\in M$ \textit{for} $1\leq i\leq m$
\textit{and} $y_1'=gy_1$,
$y_2'=gy_2$.

We shall \textit{prove} Lemma 3.2 by induction in $m$.

For $m=0$ it reduces to Proposition 1.1.
Assume that it is established
for some $m\geq 0$. We shall prove that
it is then also valid for arbitrary
$m+1$ points $x_1,...,x_{m+1}$ and
mutually non-isotropic pairs
$\f y_1, y_2\g$, $\f y_1', y_2'\g$ in $M$.
According to the assumption
there exists a $g'\in C$ such that
$g'x_i\in M$ for $1\leq i\leq m$ and
$y_1'=g'y_1$, $y_2'=g'y_2$.
If $g'x_{m+1}\in M$ we are business.
If $p \, :\, = \, g'x_{m+1}\in K_{\infty}$
then there exists
an element $h\,$, arbitrarily close
to the group unit in the stabilizer
$C_{y_1',y_2'} \, \f \, \subset C \, \g$ of the pair
$y_1'$, $y_2'$ such that $hp  \notin K_{\infty} \,$.
We prove this statement in Appendix C.
To complete the proof of Lemma 3.2 it remains to choose
$h$ so that $hg'x_i\in M$ for
$i \, = \, 1 \, , \, ... \, , \, m \,$.
This is possible since $M$
is an open set in $\M$ and $C$ acts continuously on $\M$.
Hence, $g=hg'$ satisfies the conclusion of Lemma 3.2.
\spcc

We continue with the proof of Theorem 3.1.
Assume first that $\W$ satisfies
(\textit{TS}), (\textit{L}) and
(\textit{GCI}). Lemma 3.2 and (\textit{GCI})
imply the locality
property (\textit{L}) whenever ${x_{i\ i+1}}^2\neq 0$.
Then $\W$ will be $\Z_2$-symmetric in the domain
$U \subset M^{\times n}$ of all
$\f x_1,...,x_n \g$ which are mutually non-isotropic.
Since
$\W_{1...n}$ is a (tempered) distribution its singularities
have a finite order. Therefore, there are integers $\mu^n_{ij}$
such that
\beq\label{2.8}
{\cal P}_{1...n}\f x_1,...,x_n\g\ =\
\f \mathop{\prod}\limits_{1\leq i<j\leq n}
\f {x_{ij}}^2\g^{\mu^n_{ij}} \g \ \W_{1...n}\f x_1,...,x_n\g \qquad
\eeq
is a translation invariant distribution that is
$\Z_2$-symmetric in the
entire Cartesian product space
$M^{\times n}\,$. The Fourier transform of
$P_{1...n}\f x_{12},...,x_{n-1\ n}\g\ =\ {\cal P}_{1...n}
\f x_1,...,x_n\g$,
\beq
\tilde{P}_{1...n}\f q_1,...,q_{n-1}\g\ =\
\mathop{\int \cdots \int}\limits_{M^{\times \f n-1\g}}
P_{1...n}\f y_1,...,y_{n-1}\g
\ e^{-i\f q_1y_1+...+q_{n-1}y_{n-1}\g}
\ d^Dy_1...d^Dy_{n-1} \qquad
\eeq
is obtained from
$\tilde{W}_{1...n}\f q_1,...,q_{n-1}\g$ by the
action of a differential operator in
$q_1,...,q_{n-1}$ with constant
coefficients; hence
$supp\ \tilde{P}_{1...n}\f q_1,...,q_{n-1}\g
\ \subseteq\ supp\ \tilde{W}
\subseteq {\f\overline{V}^{\ +}\g}^{\times \f n-1\g}$.
On the other hand,
the total $\Z_2$-symmetry
of ${\cal P}$ implies
\beq
{\cal P}_{1...n}\f x_1,...,x_n\g\ =\
{\cal P}_{n...1}\f x_n,...,x_1\g\quad\Rightarrow\quad
P_{1...n}\f y_1,...,y_{n-1}\g\ =\
\epsilon \ P_{n...1}\f -y_{n-1},...,-y_1\g \quad  \qquad
\eeq
where $\epsilon=\prod_i \epsilon_{i\ i+1}\ .$
Hence
$\tilde{P}_{1...n}\f q_1,...,q_{n-1}\g =
\epsilon \tilde{P}_{n...1} \f -q_{n-1},...,-q_1\g$
implying $supp\ \tilde{P}_{n...1}\subseteq
{\f \overline{V}^{\ -}\g}^{\times \f n-1\g}$.
Since the order of field labels is arbitrary and
$\overline{V}^{\ +}\bigcap\overline{V}^{\ -}=
\left\{0\right\}$
we conclude that $P\f y_1,...,y_{n-1}\g$
is a polynomial and the
same is true for ${\cal P}\f x_1,...,x_n\g$ .

If we combine this result with the fact that $W$
admits an analytic continuation
$W\f \zeta_1,...,\zeta_{n-1}\g$
in the backward tube
$\f T_-\g^{\times \f n-1\g}$ as a consequence
of energy positivity then we
end up with Eq. (\ref{2:12}).
Actually, both sides of (\ref{2:12}) are equal
in the domain $U$ introduced above
(in accord with (\ref{2.8})). On the other hand,
they have analytic continuations in the tube domain,
which thus must be equal, too.
Clearly, the
rational function ${\cal R}_{1...n}\f x_1,...,x_n\g$
so obtained is fully $\Z_2$-symmetric
and conformally invariant.

Conversely, if $\W$ is given by (\ref{2:12})
for a completely
$\Z_2$-symmetric and
conformally invariant
${\cal R}_{1...n}\f x_1,...,x_n\g$, then conditions
(\textit{TS}) and (\textit{L}) follow as
$\W$ is the boundary
value of a symmetric analytic function
in the tube domain. Condition
(\textit{GCI}) is a corollary of Proposition
2.2 since ${\cal P}_{1...n}$
is a regular multiplicator.$\quad\Box$
\spcc

{\samepage
\textit{Remark 3.2} By the proof of Theorem 3.1
it is clear that the (\textit{GCI}) condition is
just needed for extending the locality property
(\ref{Loc}) for all non-isotropic separations.
The rationality of the Wightman functions
(of the type (\ref{2:12})
without conformal invariance)
is then equivalent to this strong locality
property and energy positivity (\textit{TS}).$\quad \Box$ }



\sectionnew{
Constraints on pole degrees coming from Wightman positivity
}

Up to this point we did not use
Hilbert space (Wightman) positivity.
Taking it into account allows to deduce
that the order of poles in
correlation functions are uniformly
bounded with respect to the number
$\f n\g$ of points.

{\bf Theorem 4.1} \textit{Let}
$\phi\f x\g\ \f=\ \phi_a\f x\g\g$ \textit{be a}
(\textit{multicomponent})
\textit{field satisfying Wightman axioms
as well as} \textit{the condition}
(\textit{GCI}) \textit{for Wightman
functions}. \textit{Then the orders of the poles of the
rational functions} $\W\f x_1,...,x_n\g$
\textit{are uniformly bounded,
-i.e., the integers} $\mu^n_{jk}$
\textit{in} (\ref{2:12}) \textit{can be
chosen independent of} $n$.

\textit{Proof.} Consider the vector valued distributions
\beq\label{2.10}
\Phi_{1...k}\f x_1,...,x_k\g\ =\
\phi_1\f x_1\g...\phi_k\f x_k\g\ |0\ra \qquad
\eeq
It defines a continuous (finite
order) map of the test function space $\Ss \f M^{\times k}\g$
into the Hilbert space ${\cal H}$ (\cite{SW 78} Sec.3.3
or \cite{Jost} Sec.3.3B).
Moreover, $\Phi$ is the boundary value
of an analytic vector valued function
$\Phi_{1...k}\f x_1+iy_1,...,x_k+iy_k\g$
holomorphic for $y_k\in V^+$,
$y_{i\ i+1}\in V^-$, $i=1,...$, $k-1\,$.
For $k=2$ it follows from
Theorem 3.1 and from Reeh-Schlieder theorem
(\cite{SW 78}, Theorem 4-2)
that
$\Phi_{12}\f x_1,x_2\g =\epsilon_{12}\Phi_{21}\f x_2,x_1\g$
for any pair of
mutually non-isotropic $\f x_1,x_2\g$.
Taking into account the finite
order of the distribution $\Phi_{21}$
(in $\Ss '$) we deduce that there
is a positive integer $\mu$ such that the distributions
\beq\label{2.11}
f_{12}\f x_1,x_2\g\ =\ \f x_{12}^{\ 2}\g^{\mu}\
\la \Psi\ |\ \Phi_{12}\f x_1,x_2\g \ra
\ =\ f_{21}\f x_2,x_1\g
\quad \mathrm{for \;\; all} \quad
\Psi\in {\cal H}\;\; ,\;\;
\f x_1,x_2\g\in M^{\times 2} \qquad
\eeq
are $\Z_2$-symmetric
on $M\times M \,$.
Choosing, in particular,
$\la\Psi |=\la 0|\phi_1\f \zeta_1\g,...,\phi_k\f\zeta_k\g$
with $\left\{\zeta_j=x_j+iy_j\ ;\ 1\leq j\leq k\right\}$
in the above tube domain of analyticity and substituting
$\Phi_{12}\f x_1,x_2\g$ by
$\Phi_{k+1\ k+2}\f x_{k+1},x_{k+2}\g$
we deduce that the order of the pole of
$\W_{1...k+2}\f x_1,...,x_{k+2}\g$
in ${x_{k+1\ k+2}}^2$ does not
exceed $\mu$ and is hence
independent of $k$. Locality
then implies that this
holds for any pair of arguments.$\quad\Box$
\spcc

We proceed now to estimate
(give a realistic upper bound) of the value
of $\mu$ in (\ref{2.11}) in the important special case of
\textit{$4$ dimensional space time},
in which the ($4$-fold covering of the) conformal group
$C$ coincides with a group of $4\times 4$
pseudounitary matrices:
\beqa\label{2:20}
&& C\ =\ Spin_0 \f  4,2\g \ =\ SU \f  2,2\g \ =\
\left\{ u \in SL \f  4,\C\g \; ,\quad u\beta u^*
\ =\ \beta \
\right\} \quad , \qquad \nn
&& \mathrm{where}\quad \beta\ :=\ \f
\begin{array}{cccc}
0 & 0 & 1 & 0 \\
0 & 0 & 0 & 1 \\
1 & 0 & 0 & 0 \\
0 & 1 & 0 & 0
\end{array} \g \quad .\qquad
\eeqa

We shall
assume that our fields $\phi$
transform under
\textit{elementary induced representations} of
$C$ (see \cite{M 77} or Sec.2A of \cite{To 85}).
This means, in particular,
that $\pi_x\f g\g$ of Eq. (\ref{2|2})
provides an $x$-independent finite dimensional
\textit{irreducible representation} (IR) of the
\textit{quantum mechanical Lorentz group}
$SL \f 2,\C \g$ with
dilations:
\beq\label{2:21}
{\R}^+ \times SL \f 2,\C \g \ =\
\left\{
\f
\begin{array}{cc}
A & 0 \\
0 & {A^{*}}^{-1}
\end{array} \g
\in SU \f 2,2 \g \quad ,\quad \rho \ =\ det\ A \ >\ 0
\right\} \quad .\qquad
\eeq
(For
$\mathop{x}\limits_{\sim}=
x^0 {\bf 1} + \underline{x} \ \underline{\sigma}\ ,$
where $\sigma_1\ ,\ \sigma_2\ ,\ \sigma_3$
are the Pauli matrices we have
$\rho \mathop{\Lambda x}\limits_{\sim}=
A\mathop{x}\limits_{\sim}A^*$
where ${\f \Lambda x \g}^{\mu}=
\Lambda^{\mu}_{\ \nu}x^{\nu}$
is a proper
Lorentz transformation.) It follows that the representation
$\pi \f g_{\rho} \g$ of the
dilation subgroup
$g_{\rho} :\ x \mapsto \rho x \; \f \rho>0 \g$ is
scalar:
\beq\label{2:22}
g_{\rho}\; :\phi \f x\g \;\mapsto\;
\pi_x {\f g_{\rho}\g}^{-1} \phi \f \rho x\g =
\rho^d \phi \f \rho x\g \quad ,\qquad d>0 \quad . \qquad
\eeq
The exponent $d$ is called the (\textit{conformal})
\textit{dimension} of $\phi$.
Labeling, as customary, the IR of
$SL\f 2,\C\g$ by a pair
of (non-negative) half integers
$\f j_1,j_2\g$ ($2j_1$ and $2j_2$
giving the numbers of undotted and dotted
indices of the spin-tensor $\phi$) one proves that for
singlevalued $\pi_x \f g\g$
the (positive) number $d+j_1+j_2$ should
be an integer.
(In other words, $d$ should be an integer for Bose
fields, for which $j_1+j_2 \in \Z_+ \ ,$
and half odd integer for
Fermi fields.)

Let us note some implications
for a Wightman QFT coming from
the additional condition (\textit{GCI})
(in $4$ dimension).
First, the Wightman functions have an analytic continuation
in the domain of all mutually-nonisotropic
points of
$M_{\C}\,$. They are rational and
conformal invariant, and in
particular their euclidean restrictions,
the Euclidean Green functions,
satisfy the condition of weak conformal invariance
studied by L\"{u}scher and Mack in \cite{LM 75}. Therefore
(\cite{LM 75}, Prop.1)
there exists a unitary representation
in the Hilbert space ${\cal H}$ of
physical states, of the quantum-mechanical conformal group
$\tilde{C} \,$, the universal covering of $C\,$.
Thus (following \cite{M 77'})
${\cal H}$ can be decomposed
in a direct sum or integral of
irreducible representation spaces
of $\tilde{C} \,$.
It was shown by Mack in \cite{M 77}
that all irreducible unitary
representations
of $\tilde{C}$
with positive energy
are \textit{field representations} acting on
\beqa\label{n2:34}
{\cal H}_{\f j_1,j_2;d \g}\ =\
Span
\left\{ \ \phi^{\f j_1,j_2;d \g } \f x \g \, |0\ra \ \right\}
\quad ,\qquad
\eeqa
where $\phi^{\f j_1,j_2;d \g } \f x \g$
is a field which transforms
under the elementary induced representation of $\tilde{C}$
of weight $\f j_1,j_2;d \g \,$.
In our case, because of the rationality
and the conformal invariance
on $\M$ for the Wightman functions,
it follows that in the decomposition
of ${\cal H}$ take part only such
${\cal H}_{\f j_1,j_2;d \g}$
for which $d+j_1+j_2\,$ is an integer.
Thus the decomposition of ${\cal H}$ will be a
direct sum:
\beqa\label{n2:35}
{\cal H}\ =\ \C\, |0\ra \;\ \oplus \;\
\mathop{\oplus}\limits_{d+j_1+j_2 \in \N} \
\f {\cal N}_{\f j_1,j_2;d \g}
\otimes {\cal H}_{\f j_1,j_2;d \g} \g
\quad , \qquad
\eeqa
where the Hilbert space ${\cal N}_{\f j_1,j_2;d \g}$ gives
the multiplicity
of ${\cal H}_{\f j_1,j_2;d \g}$ in ${\cal H}\,$.

Energy positivity and unitarity restrict each dimension $d$
according to the following result:

{\samepage
{\bf Theorem 4.2} (\cite{M 77};
also see Theorem 3.18 of \cite{To 85}).
\textit{If} $\phi$
\textit{is an elementary conformal field of
weight} $\f j_1,j_2;d\g$
\textit{then requirement} (\textit{TS})
\textit{and Wightman positivity imply}
\beq\label{x3}
d\geq j_1+j_2+1 \quad \mathrm{for} \quad
j_1j_2=0 \;\; ;\quad
d\geq j_1+j_2+2 \quad \mathrm{for} \quad
j_1j_2>0 \;\; .\qquad
\eeq

The \textit{proof} is based, in part,
on an analysis of the positivity
properties of the conformally invariant $2$-point function
\beq\label{2:28}
\la 0|\phi_{j_2j_1}^* \f x \g \phi_{j_1j_2} \f y \g |0\ra
\ =\
\frac{H_{2j_1+2j_2}\f x-y \g}
{ {\left({\f x-y \g}^2 \! +
i0\f x^0-y^0 \g \right)}^{d+j_1+j_2}} \qquad
\eeq
where $H_n\f x\g$ is a
tensor-valued homogeneous harmonic polynomial of degree $n$
that is determined (up to a normalization constant) from
conformal invariance.$\quad \Box$}
\spcc

\textit{Example 4.1} If $J^{\mu}\f x \g$
is a conserved current in $D$
dimension then the $2$-point function
$\la 0|J^{\mu}\f x_1\g J_{\nu}\f x_2\g |0\ra$
is proportional to
${\f x_{12}^{\ 2}+i0x_{12}^0\g}^{-D}\f x_{12}^{\ 2}
\delta_{\nu}^{\mu}-2x_{12}^{\mu}{\f x_{12} \g}_{\nu}\g \ $.
The harmonic polynomial
$H_{2l\quad\nu_1..\nu_l}^{\quad\mu_1...\mu_l}\f x\g$
appearing in the $2$-point function
of a rank $l$ symmetric traceless
tensor is written as a symmetrized product of factors
of the type
$x_{12}^{\ 2}\delta_{\nu}^{\mu}-
2x_{12}^{\mu}{\f x_{12} \g}_{\nu}$
with subtracted traces.$\quad \Box$
\spcc

{\bf Proposition 4.3} \textit{Let a system of fields}
(\textit{in dimension} $D=4\,$)
\textit{satisfy the conditions of
Theorem} 4.1. \textit{Let}
$\phi \f x \g$ \textit{and} $\psi \f x \g$ \textit{be
two fields
in this system which are
transforming under elementary induced
representations of} $C$
\textit{of weights} $\f j_1',j_2';d' \g$ \textit{and}
$\f j_1'',j_2'' ;d'' \g$
\textit{respectively.
Then the pole degree} $\mu$ \textit{of}
${\left({\f x-y \g}^2 \! + i0\f x^0-y^0 \g \right)}^{-\mu}$
\textit{in any Wightman function}
$\la 0| ...\phi \f x \g ... \psi \f y \g ...|0\ra$
\textit{has the upper limit}:
\beqa\label{n2:?1}
\mu \ \leq \ \left[ \!\!
\left[ \ \frac{d'+j_1'+j_2'+d''+j_1''+j_2''}{2} \,-\,
\frac{1-\delta_{j_1' j_2''} \, \delta_{j_2' j_1''} \, \delta_{d'd"}}{2}
\ \right] \!\! \right]
\quad , \qquad
\eeqa
\textit{where}
$\, \left[ \! \left[ a \right] \! \right] \,$
\textit{stands for the integer part of the real number}
$a$ (\textit{i.e. the maximal}
$n \in \Z$ \textit{for which} $n \leq a\,$).

\textit{Proof.} As it was pointed out in the proof of
Theorem 4.1,
$\mu$ is the order of the vector-valued distribution
$\phi \f x \g \psi \f y \g |0\ra \,$.
Then because of the decomposition
(\ref{n2:35}) of the physical Hilbert space
${\cal H}\,$, $\mu$ does
not exceed the order in $x-y$ of the two-point function
$\la 0| \phi \f x \g \psi \f y \g |0\ra \,$,
or the maximal order of
possible (non-zero)
three-point conformal invariant Wightman functions \\
$\la 0| \phi \f x \g \psi \f y \g
\phi^{\f j_2,j_1;d \g} \f z \g |0\ra \,$.
The two-point function of $\phi$ and $\psi$ may be non-zero
only for $\f j_1',j_2' \g = \f j_2'',j_1'' \g \,$,
$d'=d''$
and then it saturates the upper limit
(\ref{n2:?1}), according to (\ref{2:28}).
It remains to verify (\ref{n2:?1})
for the orders $\mu_{\f j_1 , \, j_2 ; \, d \g}$
of the pole in $x-y$ of the three-point functions 
$\la 0| \phi \f x \g \psi \f y \g
\phi^{\f j_2,j_1;d \g} \f z \g |0\ra \,$.
We will use the results in \cite{M 77'} (Lemma 10) for the
general form
of such three-point functions. Thus we obtain:
\beqa\label{n2:?2}
\mu_{\f j_1 , \, j_2 ; \, d \g} \ \leq \ \frac{1}{2}
\left( \, d'+d''-d+L_{j_1j_2} \, \right)
\quad , \qquad
\eeqa
where $L_{j_1j_2}$ is the maximal of
the integers $l$ for which
the $SL\f 2,\C \g$ representation
$\f \frac{l}{2},\frac{l}{2} \g$
occurs in the triple tensor product
\beq\label{n2:?3-}
\f j_1',j_2' \g \otimes \f j_1'',j_2'' \g
\otimes \f j_1,j_2 \g
\eeq
(when such $l$ does not exist then
the three-point function must be zero).
Since $\f j_1'+j_1''+j_1  ,\right.$
$\left. j_2'+j_2''+j_2 \g$
is the maximal weight
occurring in the product (\ref{n2:?3-}), then
when $L_{j_1j_2}$ exists it will be equal to:
\beqa\label{n2:?4}
L_{j_1j_2} \ =\ 2 \, min \,
\left\{\ j_1'+j_1''+j_1 \ ,\ j_2'+j_2''+j_2 \ \right\}
\quad . \qquad
\eeqa
This will certainly happen for
when $j_1'+j_1''+j_1  = j_2'+j_2''+j_2\,$.
Assume, for the sake of definiteness that
\beqa\label{n2:?5}
\frac{1}{2}\ L_{j_1j_2}
\ =\ j_1'+j_1''+j_1 \ \leq \ j_2'+j_2''+j_2
\quad ; \qquad
\eeqa
then from Theorem 4.2 (Eq. (\ref{x3}))
and Eqs. (\ref{n2:?2}),
(\ref{n2:?5}) we obtain:
\beq\label{n2:?6}
\mu_{\f j_1 , \, j_2 ; \, d \g} \ \leq \ \frac{1}{2}
\f d{\,}'+d{\,}''+j_1'+j_1''+j_2'+j_2'' \g -
\frac{1+\theta_{j_1j_2}}{2}
\quad ,\qquad
\eeq
where $\theta_0=0\ ,\ \theta_{ij}=1$ for $ij>0\ $.
We further need to maximize
$-\frac{1+\theta_{j_1j_2}}{2}$ or to minimize
$\theta_{j_1j_2}\,$.
If $j_1'+j_1''+j_1  = j_2'+j_2''+j_2$ then
we can choose $\f j_1,j_2 \g = \f j,0 \g \ $or$\ \f 0,j \g$
when $\theta_{j_1j_2}=0\,$.
So we obtain the upper limit (\ref{n2:?1})
using the condition $\mu \in \Z$ (owing
to the rationality
of Wightman functions).$\quad \Box$
\spcc

{\bf Corollary 4.4} \textit{Under the assumptions of Proposition} 4.3
\textit{for} $\phi = \psi^*$ \textit{each truncated Wightman function}
$\la 0| ...\psi^* \f x \g ... \psi \f y \g ...|0\ra^T$ \textit{will have
a strictly smaller power} $\mu$ \textit{of the pole in}
${\f x-y \g}^2 \! + i0\f x^0-y^0 \g$ \textit{than the} 2-\textit{point
function} $\la 0| \psi^* \f x \g \psi \f y \g |0\ra \,$.
\textit{If the weight of} $\psi$ \textit{is} $\f j_1,j_2 \, ;d \g$ \textit{then}
\beqa\label{new4.17}
\mu \, \leq \, d + j_1 + j_2 - 1 \quad . \qquad
\eeqa

\textit{Proof.} It follows from the definition of truncated functions
(\cite{Jost} Sec.3.5C) that the \\ order of the pole in $\f x-y \g^2$ does not exceed
the order of the vector valued distribution
$\f {\bf 1}_{\cal H} - | 0 \ra \la 0 | \g \,
\psi^* \f x \g \psi \f y \g |0\ra \,$.
Hence, we only need to take into account the non-vacuum contributions
in the decomposition (\ref{n2:35}). These are determined by the 3-point
functions. The estimate (\ref{new4.17}) then follows from
Eq. (\ref{n2:?6}).$\quad \Box$
\spcc

The following example illustrates the case of $\phi \neq \psi^*$
(when the 2-point function of the fields in the product vanishes).

\textit{Example 4.2} Let $\psi$ be the free
massless Dirac field
transforming under the reducible representation
$\f \frac{1}{2},0\g \oplus \f 0,\frac{1}{2}\g$
of $SL\f 2,\C\g$
(it becomes irreducible if we extend
the Lorentz group by space
reflections). Its (normalized)
$2$-point function is given by
(\ref{2:28}) with $d+j_1+j_2=2$ and
$H_1\f x\g =\frac{1}{2\pi^2}x^{\mu}\gamma_{\mu}\ $.
The leading term of
the operator product expansion of $\psi$ with the conserved
current
$\ J^{\nu}\f x \g =\ : \!
\tilde{\psi}\f x \g \gamma^{\nu} \psi\f x \g \! : \ $
saturates the bound (\ref{n2:?1}):
{\samepage
\beqa\label{2:30}
&&\psi \f x \g J^{\nu}\f y \g |0\ra \ \simeq \
\la \psi \f x \g \tilde{\psi} \f y \g \ra
\f \psi \f y \g |0\ra + O\f {\f x-y \g }^2 \g \g \ = \nn
&& \qquad \qquad \qquad \;\;\,
=\ \frac{\gamma \f x-y \g \psi \f y \g+
O\f {\f x-y \g }^2 \g}
{2\pi^2\left[ {\f x-y \g}^2+i0\f x^0-y^0 \g \right]^2}
\ |0\ra
\quad .\qquad
\eeqa
We note that Eq. (\ref{n2:?2}) gives
a better estimate for the degree of the pole corresponding
to the contribution of the field
$\f j_1 , j_2 ; d \g$ to the operator
product expansion.$\quad \Box$}
\spcc

This example is typical in that
the leading term in the small
distance expansion of the product
of a charge carrying conformal
field with a conserved current
(or the stress energy tensor) is
the term involving the same charged field. This property is
a consequence of the Ward(-Takahashi) identity.



\sectionnew{
Cluster property. Discussion
}

The conformal hamiltonian $H$ is the
Hermitean generator of the conformal Lie algebra
corresponding to $J_{-1,0}$
(i.e., the
generator of the rotations in
the euclidean $\f -1,0 \g -$plane).
The significance of both $H$
and the associated conformal time
variable $\tau$ has been repeatedly
stressed by I. Segal
(see, e.g. \cite{Se 71}) who has pointed out,
in particular, that energy positivity with respect to the
usual relativistic energy operator $P^0$ implies  positivity
of $H$ because of the relation
\beqa\label{add.1}
H \, = \, P^0 \, + \, U_w \, P^0 \, U_w^{\, -1} \qquad
\eeqa
where $U_w$ is the representation of the Weyl group element
$w$ (of Example 2.1). Wightman axioms (including the
uniqueness of the vacuum state) together with the GCI
postulate imply that the spectrum of $H$ is contained in
$\Z_+$ (because of the rationality of Wightman functions)
and that there is a unique state in the Hilbert space
${\cal H}$ (the state space of our QFT), the vacuum,
corresponding to eigenvalue $0$ of $H \,$.
In fact, these properties of the conformal
Hamiltonian in the vacuum supeselection sector,
conversely, implies rationality of correlation
functions of local observable fields.
In terms of the
\textit{contraction semigroup}
$\left\{ e^{-tH} \, , \, t \geq 0 \right\}$ exploited
in \cite{LM 75} the cluster decomposition property can be
formulated as follows (cf. \cite{LM 75} $\mathrm{Sec}.\, 5$).

{\samepage
{\bf Proposition 5.1}
\textit{For any pair of vectors}
\beqa\label{add.2}
\la \Phi | \ = \ \la 0 | \ \phi_1 \f x_1 \g \, ... \
\phi_n \f x_n \g \quad , \quad | \Phi' \ra  \ = \
\phi'_1 \f y_1 \g \, ... \ \phi'_l \f y_l \g  \ | 0 \ra
\qquad
\eeqa
\textit{the following limiting factorization property holds}
\beqa\label{add.3}
\mathop{lim}\limits_{t \to \infty} \
\la \Phi | \ e^{-tH} \ | \Phi' \ra \ = \
\la \Phi | 0 \ra \, \la 0 | \Phi' \ra \quad . \qquad
\eeqa
$\Box$ }
\spcc

Eq. (\ref{add.3})
can be rewritten in terms of Wightman functions as
\beqa\label{add.4}
&& {\bf 1} \, \otimes \, {\bf ...} \, \otimes \, {\bf 1} \,
\otimes \, \pi_{y_1} \f e^{-tH} \g \, {\bf ...} \, \otimes \,
\pi_{y_l} \f e^{-tH} \g \ \W_{k+l}
\f x_1,...,x_k,e^{-tH}y_1,...,e^{-tH}y_l \g \
\mathop{\longrightarrow}\limits_{t \to \infty} \qquad \nn
&& \qquad \qquad \qquad \qquad \qquad
\mathop{\longrightarrow}\limits_{t \to \infty} \
\W_k \f x_1,...,x_k \g  \ \W_l \f y_1,...,y_l \g
\quad . \qquad
\eeqa
The limit in Eq. (\ref{add.4}) may be
given two different (valid) interpretations:
first, as a limit for every fixed set of
points $\f x_1,...,x_k \g$ and $\f y_1,...,y_l \g$ in
the corresponding tube domains; second, as a limit of
rational functions (because of the rationality of both sides
of (\ref{add.4})).
Note that $\left\{ e^{-tH} \ , \ t \geq 0 \right\}$ is a
subsemigroup of the complex conformal group $C_{\C}\,$.
Theorem 3.1 implies that in a QFT satisfying GCI the
correlation functions viewed as rational functions would
satisfy (\ref{add.4}) for any semigroup
$h \, e^{-tH} \, h^{-1}$
conjugate to $e^{-tH}$ in $C_{\C}\,$.
This is true, in particular, for the
two opposite dilation
semigroups:
\beqa\label{add.5}
U_{\rho} \, \phi'_i \f y_i \g \, U_{\rho}^{-1} \ = \
\rho^{d'_i} \, \phi'_i \f \rho y_i \g \quad , \quad
i \, = \, 1 \, , \, ... \, , \, l \quad , \quad
\mathrm{for} \quad \rho \, \geq \, 1 \quad
\mathrm{or} \quad \rho \, \leq \, 1 \quad . \qquad
\eeqa
As a corollary of Proposition 5.1 we have:
\beqa\label{add.6}
\rho^{d'_1+...+d'_l} \,
\W_{k+l} \f x_1,...,x_k,\rho y_1,...,\rho y_l \g \
\mathop{\longrightarrow}\limits_{\rho^{\pm 1} \, \to \, 0} \
\W_k \f x_1,...,x_k \g  \ \W_l \f y_1,...,y_l \g
\quad . \qquad
\eeqa
Combined with locality this gives (\cite{Jost} Sec.3.5C)

{\samepage
{\bf Proposition 5.2}
\textit{For any splitting of the arguments}
$\f x_1,...,x_n \g$ \textit{of a truncated Wightman function}
$\W_n^{\mathrm{T}}$ \textit{into two disjoint subsets}
$x_{i_1},...,x_{i_k}$ \textit{and} $x_{j_1},...,x_{j_l} \;$,
$k+l \, = \, n \;$,
\textit{we have}
\beqa\label{add.7}
\rho^{d_{j_1}+...+d_{j_l}} \,
\W_n^{\mathrm{T}}
\f x_{i_1},...,x_{i_k},\rho x_{j_1},...,\rho x_{j_l}  \g
\ \longrightarrow \ 0 \quad \mathrm{for} \quad
\rho^{\pm 1} \ \to \ 0 \quad . \qquad
\eeqa
$\Box$ }
\spcc

The general principles of QFT together with GCI are so
restrictive that they allow, in principle, the computation of
conformally invariant correlation functions. We shall
illustrate this fact by writing down the general $4$-point
function of a neutral scalar field $\phi$ of low conformal
dimension.

{\bf Proposition 5.3} \textit{The general}
$4$-\textit{point function of a neutral scalar
field} $\phi$ \textit{of (an integer) dimension}
$d$ \textit{satisfying} (\textit{TS}) (\textit{L})
(\textit{GCI}) \textit{and the implications of
positivity contained in Proposition}$\, 4.3$
\textit{has the form}
\beqa\label{5.8}
&&
\W \f x_1,x_2,x_3,x_4 \g \ = \
{\cal D}_d \f x_1,x_2,x_3,x_4 \g \,
{\cal P} \f \eta_1 , \eta_2 \g
\quad , \quad \nn
&&
{\cal D}_d \f x_1,x_2,x_3,x_4 \g \ = \
{ \f \frac{x_{13}^{\;\  2} \  x_{24}^{\;\  2} }
{\f x_{12}^{\;\  2} + i0x_{12}^0 \g
\f x_{23}^{\;\  2} + i0x_{23}^0 \g
\f x_{34}^{\;\  2} + i0x_{34}^0 \g
\f x_{14}^{\;\  2} + i0x_{14}^0 \g} \g }^d
\quad , \qquad \qquad
\eeqa
\textit{where} $\eta_{1,2}$ \textit{are the cross-ratios}
\beqa\label{5.9}
\eta_1 \ = \
\frac{x_{12}^{\;\  2} \  x_{34}^{\;\  2}}
{\f x_{13}^{\;\  2} + i0x_{13}^0 \g
\f x_{24}^{\;\  2} + i0x_{24}^0 \g}
\quad , \quad
\eta_2 \ = \
\frac{x_{14}^{\;\  2} \  x_{23}^{\;\  2}}
{\f x_{13}^{\;\  2} + i0x_{13}^0 \g
\f x_{24}^{\;\  2} + i0x_{24}^0 \g}
\quad , \qquad \qquad
\eeqa
${\cal P}$ \textit{is a polynomial in}
$\eta_1 \, , \, \eta_2$ \textit{of overall degree} $2d \,$,
\beqa\label{5.10}
{\cal P} \f \eta_1 \, , \, \eta_2 \g \ = \
\mathop{\sum}
\limits_{\mathop{}
\limits^{\mu \, \geq \, 0 \; , \; \nu \, \geq \, 0}_{\mu
\, + \, \nu \, \leq \, 2d }}
\; \;
C_{\mu \nu} \ \eta_1^{\, \mu} \, \eta_2^{\, \nu}
\quad . \qquad
\eeqa
\textit{Locality implies invariance of} ${\cal P}$
\textit{under the} $6$ \textit{element dihedral group}
${\cal D}_3 \f \cong {\cal S}_3 \g \,$:
\beqa\label{5.11}
s_{12} && \;\; : \qquad \eta_2^{\, 2d} \
{\cal P} \f \frac{\eta_1}{\eta_2} \, , \,
\frac{1}{\eta_2} \g \ = \
{\cal P} \f \eta_1 \, , \, \eta_2 \g
\qquad \nn
s_{12} && \;\; : \qquad \eta_1^{\, 2d} \
{\cal P} \f \frac{1}{\eta_1}  \, , \,
\frac{\eta_2}{\eta_1} \g \ = \
{\cal P} \f \eta_1 \, , \, \eta_2 \g
\qquad \nn
s_{13} \, = \, s_{12} \, s_{23} \, s_{12}
\, = \, s_{23} \, s_{12} \, s_{23} && \;\; : \qquad
{\cal P} \f \eta_2 \, , \, \eta_1 \g \ = \
{\cal P} \f \eta_1 \, , \, \eta_2 \g
\qquad \qquad \qquad \qquad \qquad \qquad
\eeqa
\textit{($s_{ij}$ standing for the permutation of the
arguments $i \, , \, j$ of the rational function $\W \,$).
The normalization of} ${\cal P}$ \textit{is related to
the normalization of the} $2$-\textit{point function of}
$\phi \,$:
\beqa\label{5.12}
\W \f x_1,x_2 \g \ = \
\frac{N_d}{{\f x_{12}^{\;\  2} + i0x_{12}^0 \g }^d}
\quad \Longrightarrow \quad
{\cal P} \f 1,0 \g \ = \  N_d^{\, 2} \  = \
{\cal P} \f 0,1 \g
\quad . \qquad
\eeqa
\textit{The truncated} $4$-\textit{point function admits
a similar representation,}
\beqa\label{5.13}
\W_T \f x_1,x_2,x_3,x_4 \g \ = \
{\cal D}_d \f x_1,x_2,x_3,x_4 \g \,
{\cal P}_T \f \eta_1 , \eta_2 \g
\quad , \quad
\eeqa
\textit{where the polynomial} ${\cal P}_{T}$
\textit{has the properties} (\ref{5.10}) \textit{and}
(\ref{5.11}) \textit{of} ${\cal P}$ \textit{while
instead of} (\ref{5.12}) \textit{it satisfies}
\beqa\label{5.14} {\cal P}_T \f 1,0 \g \  = \
{\cal P}_T \f 0,1 \g \ = \  0
\quad . \qquad
\eeqa

{\samepage
\textit{Proof.} The general form (\ref{5.8}) is
implied by the fact that the cross ratios
(\ref{5.9}) form
a basis of (rational)
invariants of $4$ points. The summation limits
in (\ref{5.10}) follow from Proposition 4.3.
The $i0x_{jk}^0$ prescription in
${\f x_{jk}^{\;\  2} + i0x_{jk}^0 \g }^{\, -\lambda}$
reflects energy
positivity. Eqs. (\ref{5.12}) and (\ref{5.14})
are consequences of
Eqs. (\ref{add.6}) and (\ref{add.7}),
respectively.$\quad \Box$ }
\spcc

The simplest special cases, $d = 1 \, , \ 2 \,$,
show that the range of summation in (\ref{5.10})
can be further restricted by combining the operator
product expansion implied by (\ref{5.8}) with
positivity.
Indeed, the most general polynomials ${\cal P}$
satisfying (\ref{5.11}) and (\ref{5.12})
for the above values of $d$ are
\beqa\label{5.15}
{\cal P}_1 \f \eta_1 \, , \, \eta_2 \g
&& \!\!\!\!\!\!\! = \;
C_1 \left[ \,
{\f 1 - \eta_1 - \eta_2 \g }^2
\, - \, 4 \, \eta_1 \eta_2 \, \right] \ + \
N_1^{\, 2}
\f \eta_1 + \eta_2 + \eta_1 \eta_2\g
\\
\label{5.16}
{\cal P}_2 \f \eta_1 \, , \, \eta_2 \g
&& \!\!\!\!\!\!\! = \;
C_{20} \left[ \,
{\f 1 - \eta_1^{\, 2} - \eta_2^{\, 2} \g }^2 \, - \,
4 \, \eta_1^{\, 2} \eta_2^{\, 2} \, \right]
\ + \
\nn
&& \!\! +  \
C_{21} \left[ \, \eta_1 {\f 1 - \eta_1 \g }^2 +
\eta_2 {\f 1 - \eta_2 \g }^2 +
\eta_1 \eta_2 { \f \eta_1 - \eta_2 \g }^{\, 2}
\, \right] \ + \
\nn
&& \!\! +  \
N_2^{\, 2}
\f  \eta_1^{\, 2} + \eta_2^{\, 2} +
\eta_1^{\, 2} \eta_2^{\, 2} \g \ + \
C_2 \ \eta_1 \eta_2 \f 1 + \eta_1 + \eta_2 \g
\quad . \qquad
\eeqa

On the other hand,
for $d=1$ the $2$-point Wightman
function $\W \f x_1 , x_2 \g = W_1 \f x_{12} \g$
satisfies the d'Alembert equation
$\Box \, W_1 \f x \g = 0 \,$. It then follows from
the above cited Reeh-Schlieder theorem and from
Wightman positivity that
$\Box \, \phi \f x \g = 0 \,$. Consequently,
$C_1 = 0$ in (\ref{5.15}) and we end up with a
free field theory.
More generally, the vanishing of $C_1 \,$-
as well as that of $C_{20}$ and $C_{21} \,$-
follows from Corollary 4.4.
(We owe this remark to Yassen Stanev.)

{\samepage
A systematic study of the implications of
operator product expansions combined with
positivity is relegated to the sequel
of this paper (announced in the Introduction).
}



\vspace{0.4in}
\noindent
{\bf\Large Acknowledgements} \\

We thank Yassen Stanev
for his inquisitive questioning and for
helpful discussions.
The authors acknowledge partial support by the
Bulgarian National Council for Scientific Research under
contract $F \! \! - \! 828 \ $.




\appendiX[Appendix~]
\link{equation}{section}\toheight1


\sectionnew{
The Klein-Dirac quadric
}
\label{app:A}

Viewed
as a homogeneous space of the group $SO_0 \f D,2 \g \,$,
compactified $D$-dimensional Minkowski space $\M$ can be
defined as a \textit{projective isotropic cone} (or quadric)
of signature $(D,2)$ (see \cite{Di 36}).
Points in $\M$ are identified with isotropic
rays in $\R^{D,2} \,$,
two collinear (isotropic)
vectors, $\vec\xi$ and $\l\vec\xi$ ($\l \neq 0$)
representing the same point in $\M$:
\beq\label{quadrica}
\M=Q/\R^{*} \quad , \quad
Q:=\left\{ \vec{\xi} \in \R^{D,2} \; ; \;
\vec{\xi}\neq 0, \; \vec{\xi}^2:=
\underline{\xi}^2 +\xi^2_D-\xi_0^2-
\xi^2_{-1}=0     \right\} \quad . \qquad
\eeq
Here $\R^{*}$ is the multiplicative group of non-zero reals,
$\underline{\xi}^2:=\xi^2_1+...+\xi_{D-1}^2$.
In this representation of $\M \,$,
the embedding of Minkowski space
$M$ in $\M$ is given by
the Klein-Dirac compactification
formulae:
\beqa\label{1.6}
&&
M \; \ni \; x \quad \longmapsto \quad
\left\{ \lambda \vec{\xi}_x  \right\} \; \in \; \M
\qquad , \quad \mathrm{where:} \qquad
\nn &&
\vec{\xi}_x\ = \ x^{\mu}\vec{e}_{\mu} \, + \,
\frac{1+x^2}{2} \, \vec{e}_{-1} \, + \,
\frac{1-x^2}{2} \, \vec{e}_D
\qquad
\eeqa
and
$\vec{e}_1 \, , \, ... \, , \,
\vec{e}_D \, , \, \vec{e}_{-1}
\, , \, \vec{e}_0$
is the standard
basis in $\R^{D,2}$. Then we have
the following expression for
the pseudo-euclidean interval:
\beqa\label{1.7}
(x-y)^2=-2\x_x.\x_y={\f \x_x-\x_y \g}^2 \quad .
\eeqa

Eq. (\ref{1.7}) shows
that two points $p_1,p_2 \in \M$
are (mutually) \textit{isotropic} if the inner product
$\vec{\xi}_1.\vec{\xi}_2$ of (any of) their representatives
$\vec{\xi}_1$, $\vec{\xi}_2$ is zero. Thus it is obviously a
$SO_0 \f D,2 \g$-invariant relation in
$\M \times \M \,$. The point
$p_{\infty} \in \M$ can be defined as
$p_{\infty} =
\left\{ \lambda \vec{\xi}_{\infty} \right\} \,$,
where $\vec{\xi}_{\infty} = \vec{e}_D - \vec{e}_{-1} \,$.
Then the representatives $\vec{\xi}_x$
of the points $x$ of Minkowski
space $M$ are characterized by the "normalization"
$\vec{\xi}_x.\vec{\xi}_{\infty} = 1$ so that the set
$K_{\infty}$ of points at infinity is indeed
\beqa\label{K-inf}
&&\left(\; K_{p_{\infty}} \equiv\; \right)
K_{\infty}\;= \nn &&= \left\{ p \in \M \;\; ; \,
(p,p_{\infty})=0, \; \mathrm{i.e.}, \,
\vec{\xi}.\vec{\xi_{\infty}}=0
\;\, \mathrm{for} \; p=\left\{\l\vec{\xi}\right\},
p_{\infty}=\left\{\l\vec{\xi_{\infty}}\right\}   \right\}
\quad . \qquad\;\
\eeqa
As an illustration of the transitivity
property used in the proof
of Proposition 1.1 of
Sec. 1 we note that the rotation on $\pi$ in the
$(\xi^{-1},\xi^0)$-plane interchanges the origin in $M$ with
$p_{\infty}$
(this is the Weyl reflection used in
the Examples 2.1 and 2.2).

Equation (\ref{1.7}) also allows
to compute the conformal factor
$\omega\f x,g \g\omega\f y,g \g$ multiplying the interval
${\f x-y\g}^2$ under the action of $g \in SO_0 \f D,2 \g \,$.
Indeed,
$\omega\f x, g \g$ can be defined by:
\beq\label{1:8}
g\x_x\ =\ \omega\f x,g \g \x_{gx} \quad \mathrm{for}
\quad \x_x .\x_{\infty}\ =\ 1\ =\ \x_{gx}.\x_{\infty} \quad
\Rightarrow \quad \omega\f x,g \g=g\x_x .\x_{\infty}\;
\qquad
\eeq
and is, hence, a second degree polynomial in $x \,$.
Here $\x \mapsto g\x$ is the
\textit{linear} action of the group
$SO_0 \f D,2 \g$ on $\R^{D,2} \,$,
while $x \mapsto gx$ is the
\textit{nonlinear} action of
$SO_0 \f D,2 \g$ on $M\,$ which can be computed from
(\ref{1:8}).
We then find (cf. (\ref{1:1})):
\beq\label{1:9a}
{\f gx-gy\g}^2\ =\ -2\ \x_{gx}.\x_{gy}\ =
\ -2\ \frac{g\x_x .g\x_{y}}{\omega\f x,g \g\omega\f y,g \g}
\ =\ \frac{{\f x-y\g}^2}{\omega\f x,g \g\omega\f y,g \g}
\; .\qquad
\eeq
So, the elements of the group $SO_0 \f D,2 \g$ 
act indeed conformally
on $M$ \textit{outside their 
singularities}. Because the dimension
of $SO_0 \f D,2 \g$ is equal to
that of the conformal Lie algebra
coming from
Liouville theorem (see, e.g., \cite{To 85} Appendix A)
then $SO_0 \f D,2 \g$
must be locally isomorphic to the conformal group $C_0 \,$.

\textit{The double cover} $\MM$ of $\M$
(\ref{quadrica}) is diffeomorphic
\textit{to the product of the}
$(D-1)$\textit{-sphere}, $\Ss^{D-1}$,
\textit{with the unit circle} $\Ss^1$;
$\M$ \textit{is obtained from it by
identifying opposite points}:
\beqa\label{m2}
&&\MM\,=\,\Ss^{D-1}\times\Ss^1\quad , \nn
&&\qquad\;\; \Ss^{D-1}\,=\,
\left\{\left( \underline{\xi},\xi_D\right) \; ; \;
\underline{\xi}^2+\xi_D^2=1   \right\} \quad ,\quad
\Ss^1\,=\,
\left\{\left(\xi_{-1},\xi_0\right) \; ; \xi_{-1}^2+\xi_0^2=1
\right\} \quad , \quad \nn
&&\M\ =\ {\MM} \left/ {\Z_2} \right. \ =\
{\MM} \left/ {\f \vec{\xi}\cong -\vec{\xi} \g } \right.
\quad
\left(\vec{\xi} \in \MM \right). \qquad
\eeqa

{\bf Observation:} \textit{It follows from} (\ref{m2})
\textit{that}
\textit{compactified Minkowski space} $\M$
\textit{is orientable for even} $D$
\textit{and non-orientable for odd} $D\,$.$\quad\Box$
\spcc

Eq. (\ref{quadrica}) admits a complex version,
$\M_{\C}=Q_{\C}/\C^*$
where $Q_{\C}$ is the complexification of
the quadric $Q$ and $\C^*$
is the multiplicative group of
(non-zero) complex numbers. $\M_{\C}$
is a homogeneous space of
the complexified conformal group $C_{\C}$.
$\M_{\C}$ and $C_{\C}$ admit a unique
antianalytic involution $*$
whose fixed points belong to
$\M$ and $C$, respectively. In the
above realization we have $p^*=\left\{\l\x^*\right\}$,
$\left(gp\right)^*=g^*p^*$ where $\x^*$
is the (componentiwise)
complex conjugate of $\x$ and $g^*$
is the complex conjugate of the
matrix $g\in SO_0\left(D,2;\C\right)$.



\sectionnew{
Proof of Proposition 2.2
}
\label{app:B}

We shall prove the statement by
induction in the number of points.
For $n=2$ the $2$-point vacuum correlator
is a boundary value of the
analytic function
$\left[ {\f x-\zeta \g }^2 \right]^{-\mu}$ holomorphic
for $\f x,\zeta \g \in M \times T_+$
which satisfies (\textit{GCI}) as a
consequence of (\ref{1:1}) and of
the conformal invariance of the forward
tube. (In fact, this case can also
be covered by our induction inference if
we interpret the (\textit{GCI}) property for
$n=1$ as the identity $1=1\ $.)
Assume now that the statement is established for all
$\f n-1 \g$-point functions for some $n>1\ $. To prove it
for the $n$-point functions we fix a
$g \in C$ and $n$ open sets
$U_1,...,U_n$ in $M$ whose closures
$\overline{U_1},...,\overline{U_n}$
are compact and such that the mapping $g:\ x \mapsto gx$
has no singularity for $x \in \overline{U_i}\ $.
So we will establish that Eq.(\ref{2:11})
is locally valid for
$\f x_1,...,x_n \g \in U_1 \times ... \times U_n \,$.
The distribution
in the right hand side of (\ref{2:11}) is then the boundary
value for $Im \f \zeta \g \to 0\ , \ \zeta \in T_+$
of the analytic function
\beq\label{a1}
{\W}_n \f x_1,...,x_{n-1},\zeta \g \  =\
{\W}_{n-1} \f x_1,...,x_{n-1} \g
\mathop{\prod}\limits_{j=1}^{n-1}
\left( {\f x_j-\zeta \g }^2 \right)^{-\mu_{jn}}
\quad \f \zeta \in T_+ \g \qquad
\eeq
(with ${\W}_{n-1}$ again given by the right hand side of
(\ref{2:11}) for $n$ substituted by $n-1\ $).
Applying the induction assumption to ${\W}_{n-1}$
and Eq. (\ref{1:1}) to the other factors in the product
in the right hand side of (\ref{a1})
(treating them as a multiplicator) we deduce
\beqa\label{a2}
\W_n \f x_1,...,x_{n-1},x_n+iy \g \ = &&
{\omega \f x_1 , g \g }^{-\mu_1} \, ...\,
{\omega \f x_{n-1} , g \g }^{-\mu_{n-1}}
\, {\omega \f x_n+iy , g \g }^{-\mu_n}    \nn
&& {\W}_{n-1} \f gx_1,...,gx_{n-1} \g
\mathop{\prod}\limits_{j=1}^{n-1}
\left( {\f gx_j-gx_n-
\zeta_y \f gx_n \g \g }^2 \right)^{-\mu_{jn}}
\quad ,\qquad \qquad
\eeqa
where
$\zeta_y \f gx_n \g := g\f x_n +iy \g -gx_n \in T_+$
because of
the conformal invariance of $T_+\ $.
We also have the limit:
\beqa\label{a3}
\mathop{lim} \limits_{V_+ \ni \ y \to 0} &&
{\W}_{n-1} \f x'_1,...,x'_{n-1} \g
\mathop{\prod}\limits_{j=1}^{n-1}
\left( {\f x'_j-x'_{n} -
\zeta_y \f x'_n \g \g }^2 \right)^{-\mu_{jn}} \ = \nn
&&=\ {\W}_{n-1} \f x'_1,...,x'_{n-1} \g
\mathop{\prod}\limits_{j=1}^{n-1}
\left( {\f x'_j-x'_n \g }^2 +
i0\f {x'_j}^0-{x'_n}^0 \g \right)^{-\mu_{jn}}
\quad ,\qquad
\eeqa
which combined with (\ref{a2})
gives (\ref{2:11}) (due to the
continuity of multiplication of
a distribution with a multiplicator).$\quad \Box$



\sectionnew{
Completion of the proof of Lemma 3.2
}
\label{app:C}

It remains to prove the following statement: \\
\textit{If the points} $y_1' , y_2' \in M$ \textit{are
non-isotropic,} $p \in K_{\infty}$ \textit{and}
$C_{y_1',y_2'}$
\textit{is the stabilizer of the pair} $y_1',y_2'$
\textit{in}
$SO_0 \f D,2 \g \,$,
\textit{then in any neighbourhood of unity
in} $C_{y_1',y_2'}$ \textit{there exists an element} $h$
\textit{such that}
$hp \notin K_{\infty} \,$. \\
In the case $D=1 \,$, $K_{\infty}$ consists of a single point,
$p_{\infty} \,$, and it is not stable for $C_{y_1',y_2'}$
so one does not need the argument below.
To prove
the above statement for $D \geq 2$ we shall use the representatives
$\x_1$, $\x_2$ and $\x$ of the points
$y_1'$, $y_2'$ and $p$ (cf. (\ref{1.6})) in the quadric
(\ref{quadrica}). We have
$\x_1.\x_2\neq 0\neq \x_{\infty}.\x_a$ for
$a=1,2$; $\x.\x_{\infty}=0\ \f
\Leftarrow\ \x \in K_{\infty}\g$.
The metric in $Span\left\{\x_1,\x_2\right\}$
being non-degenerate
(since $\x_1.\x_2\neq 0$ while
${\x_1}^{\ 2}={\x_2}^{\ 2}=0$) there is a
pseudo-orthogonal decomposition
\beq
\R^{D,2}\ =\ Span\left\{\x_1,\x_2\right\} \oplus
{\left\{\x_1,\x_2\right\}}^{\perp}\ .
\eeq
Let $\x_{\infty}=\x_{\infty}^{\ '}+\x_{\infty}^{\ ''}$,
$\x =\x^{\ '}+\x^{\ ''}$ $\f \x_{\f \infty\g}^{\ '}\in
Span\left\{ \x_1, \x_2\right\}\g$
be the corresponding decompositions of
$\x_{\infty}$ and $\x$.
Since
$\x_{\infty}^{\ '}.\x_{1,2}=\x_{\infty} .\x_{1,2}\neq 0$
and $\x_{\infty}^{\ 2} =0$ then
${\f\x^{\ '}_{\infty}\g}^2 \neq 0$
and ${\f\x^{\ ''}_{\infty}\g }^2 \neq 0$.
It follows also that
$\x^{\ ''} \neq 0$ (i.e. $\x$ does not belong to
$Span\left\{\x_1,\x_2\right\}$ since ${\x}^{\ 2}=0$
and $\x .\x_{\infty}=0 \,$).
Let $G$ be the connected
component of the orthogonal group of the subspace
$\left\{\x_1,\x_2\right\}^{\perp}$
$\f G\cong SO_o \f D-1,1\g \g$.
Clearly, $G$ is a subgroup of
the stabilizer of the points
$y'_i=\left\{\l\x^{\ '}_i\right\}$, $i=1,2$:
\[
G\ \subset\ C_{y_1',y_2'}
\ =\ C_{Span\left\{\x_1,\x_2\right\}}
\  : \, = \
SO_o \f {Span\left\{\x_1,\x_2\right\}}\g\ \times\
SO_o \f{\left\{\x_1,\x_2\right\}}^{\perp}\g\;.
\]
For $h$ varying in $G$, $h\x_{\infty}^{\ ''}$
moves on a hyperboloid
in $\left\{\x_1,\x_2\right\}^{\perp}$.
Consequently, the (real valued)
function
$f\f h\g =h\x . \x_{\infty}=
\x^{\ '}.\x_{\infty}^{\ '}+\x^{\ ''}.
h^{-1}\x_{\infty}^{\ ''}$
cannot be a constant.
On the other hand it is real analytic (in fact,
algebraic) and $f\f 1\g=0$.
Hence, in any (arbitrarily small)
neighbourhood of the group unit in $G$
(and also in $C_{y_1',y_2'}$)
there exist elements $h$ such that
$f\f h\g\neq 0 \,$, i.e.
$h p \notin K_{\infty}\,$.$\quad\Box$






\end{document}